\documentclass[preprint,3p,times]{elsarticle}

\usepackage{hyperref}
 \usepackage{graphicx}
\usepackage{amssymb}
\usepackage{amsthm}
\usepackage{amsmath}
\usepackage{natbib}

\usepackage[nodots,nocompress]{numcompress}

\biboptions{}

\begin{document}
\makeatletter
\def\ps@pprintTitle{%
 \let\@oddhead\@empty
 \let\@evenhead\@empty
}
\makeatother

\begin{frontmatter}


\title{In pursuit of the low-energy Solar neutron flux}


%\author[rvt]{Prithish ~Halder\corref{cor1}}
\author[rvt]{\href{https://orcid.org/0000-0002-1073-1419}{Prithish Halder\hspace{0.8mm}\includegraphics[scale=0.06]{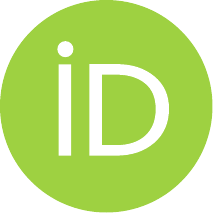}}\corref{cor1}}

\ead{dr.prithishhalder@gmail.com,phalder2@unl.edu}
%\author[rvt]{Peter A. Dowben}
\author[rvt]{\href{https://orcid.org/0000-0002-2198-4710}{Peter A. Dowben\hspace{0.8mm}\includegraphics[scale=0.06]{orcid.pdf}}}
\ead{pdowben1@unl.edu}

\address[rvt]{Department of Physics and Astronomy, University of Nebraska-Lincoln, Jorgensen Hall, Lincoln, NE 68588, USA}
\cortext[cor1]{Corresponding author}

\begin{abstract}
Understanding the origin of low-energy solar neutrons flux is crucial for probing solar energetic processes and neutron transport mechanisms in interplanetary space. This study investigates the role of coronal mass ejections (CMEs) in modulating the low-energy solar neutrons. The neutron flux is modeled by incorporating a CME-width scaling factor into existing neutron fluence models. Our analysis, based on CME data from the SECHHI COR2 database during the DANSON experiment (2016-10-27 to 2017-03-17), identified narrow-width (20$^\circ$ $<$ $\alpha$ $<$ 80$^\circ$) and fast (v $>$ 800 km/s) CMEs as key contributors to neutron production. The revised model predicts a neutron flux of 5 - 49 neutrons cm$^{-2}$ s$^{-1}$ at 1 AU for a neutron travel time of 1.66 hours, aligning with previous reports. Additionally, the estimated total flux of 2.45 MeV neutrons over a 4-hour period accounts for 10.23\% of DANSON's total flux. These findings suggest that CME-driven mechanisms may significantly influence the low-energy solar neutron flux. More observational efforts are needed to refine neutron flux estimates and improve background subtraction techniques for spaceborne neutron detectors.

\end{abstract}


\end{frontmatter}

% \linenumbers
\section{Introduction}
     \label{S-Introduction}

The presence of solar neutrons penetrating the near-Earth environment has long been theorized \citep{Biermann1951NeutraleSonne}, yet required decades of experimental effort before firm detections were achieved \citep{Chupp1982A21., Forrest1982Evidence7}. Solar neutrons are thought to arise from nuclear interactions between flare-accelerated ions and the solar atmosphere, making them valuable diagnostics of ion acceleration processes near the Sun. However, detecting solar neutrons, particularly at low energies and at 1 AU, remains difficult due to the uncharged, short-lived nature of neutrons and the dominance of background secondaries. One notable attempt was the Neutron Spectrometer (NS) aboard the MESSENGER spacecraft, which reported detections of solar neutrons in the 0.5–7.5 MeV range at 0.48 AU. However, those results were challenged due to the instrument’s inability to distinguish between primary solar neutrons and secondary neutrons generated by energetic particles interacting with the spacecraft itself. This longstanding ambiguity has limited our understanding of low-energy neutron production and survival in the heliosphere.

The DANSON experiment (Directional and Spectral Neutron Observer), deployed on the zenith-facing side of the International Space Station, was designed to address this limitation by using stacked and moderated lithium tetraborate (Li\textsubscript{2}B\textsubscript{4}O\textsubscript{7}) single crystals and boron-enriched control materials to separate neutron-induced damage from proton-induced effects \citep{Benker2019PossibleMaterials}. Isotope-enriched elements ($^{10}$B and $^{11}$B) enabled in-situ subtraction of the proton component, while placement on the zenith-facing surface minimized albedo and spacecraft-induced secondary neutron contamination. Monte Carlo simulations using MCNP6 and GEANT4 validated neutron transport and energy deposition profiles, demonstrating that the lithium tetraborate (LTB) thermoluminescence (TL) signal, in an effective neutron calorimeter, corresponded more closely to 2 to 4 MeV neutrons arriving from the solar direction.
Unexpectedly, DANSON detected a persistent, directional neutron flux in this low-energy range over a five-month period (October 2016–March 2017), with a cumulative fluence exceeding $3 \times 10^9$ neutrons/cm². No class X solar flares or strong Solar Energetic Particle (SEP) events were observed during this interval. The measured fluence exceeds existing theoretical predictions of solar neutron survival at 1 AU, particularly given that a 1 MeV neutron, with a velocity of $1.38 \times 10^7$ m/s, has only a modest probability of reaching Earth before decaying. Prior neutron production models based on proton and $^4$He reactions with heavy nuclei also suggest that such low-energy neutrons should be significantly attenuated over 1 AU \citep{Murphy2012THEHELIOSPHERE}. While earlier studies have explored the potential role of $^3$He-rich SEP events in enhancing low-energy neutron yields \citep{Murphy2016EVIDENCESPECTRUM, Murphy2017Neutron3He}, such reactions typically require unusually high $^3$He/$^4$He ratios ($\gtrsim 1$) to contribute significantly to total neutron production. It must be noted that, unlike the DANSON neutron calorimeter, the flux measurements in some earlier experiments had inadequate background subtraction, meaning the reported fluences likely represent upper bounds on the true solar neutron flux. For more common impulsive events with $^3$He/$^4$He $<$ 1, the contribution of $^3$He-induced reactions to the total neutron budget is expected to be modest. However, these reactions produce neutron energy spectra that peak around 2 MeV, making them spectrally relevant to DANSON, even if they are not energetically sufficient to explain the entire observed flux.

Given the apparent absence of traditional high-fluence neutron-producing flares or SEP events, we explored whether persistent solar magnetic activity, particularly the occurrence of numerous narrow CMEs and C-class flares, may serve as an indirect proxy for conditions conducive to low-energy neutron production. Narrow CMEs are often associated with compact reconnection events or jet-like structures that may locally accelerate ions in confined magnetic environments. Such settings, though weak in radiative signatures, could contribute to neutron production via loop-top trapping and nuclear interactions.

In this study, we first confirm the external, directional, and neutron-specific nature of the DANSON signal and rule out instrumental and ISS-based secondary sources. We then investigated whether the rate and width of CMEs during the DANSON operational period can be empirically correlated with the observed neutron accumulation, using a revised version of a probabilistic model originally developed for flare-associated neutron events. Although we do not claim that CME activity directly drives neutron generation, we explore its potential as a proxy for unresolved solar processes that may underlie the observed low-energy neutron flux.

\section{Evaluating the Origin of the Detected Neutrons}
%\textbf{The DANSON detector was installed on the zenith-facing surface of the International Space Station (ISS), with its nadir and lateral sides shielded by stowage and structural materials. The measured thermoluminescence (TL) signal exhibited a pronounced asymmetry, with the highest neutron capture signature recorded in the topmost (Layer 1) and second (Layer 2) detectors. Layer 5 facing the nadir recorded a significantly lower TL signal, particularly in the natural boron (LTB-N) detectors, resulting in a nadir-to-zenith ratio as low as ~0.36. This spatial asymmetry is inconsistent with the nearly isotropic distribution of cosmic ray-induced or spacecraft-generated neutrons measured by ISS-RAD (Zeitlin et al., 2023), where nadir-to-zenith ratios are reported to be close to unity across all modules.}

The DANSON detector was installed on the zenith-facing surface of the International Space Station (ISS), with neutron shielding on the lateral sides \citep{Benker2019PossibleMaterials}. The measured thermoluminescence (TL) signal exhibited a pronounced asymmetry, with the highest neutron capture signature recorded in the zenith-facing layers (Layer 1 \& 2) detectors. Layer 5, facing the nadir, recorded a significantly lower TL signal, particularly in the natural boron (LTB-N) detectors, resulting in a nadir-to-zenith flux ratio as low as 0.346 (with partial background) and 0.18 (with complete background subtraction). This spatial asymmetry highlights the strong influence of local shielding, and although the ISS-RAD Fast Neutron Detector \citep{Zeitlin2023ResultsDetector} is designed to respond nearly isotropically, its measurements also reveal location- and direction-dependent neutron fluxes due to complex ISS shielding geometry.

Monte Carlo simulations using MCNP6 and GEANT4 to match the observed LTB TL signal profile as a function of the detector depth. The modeled data indicated incident neutron energies between 1 and 2 MeV. Simulations for neutron energies below 0.5 MeV or above 10 MeV were not good fits to the experimental data. This narrow spectral response of the DANSON experiment is in agreement with the detection of 0.5–7.5 MeV solar neutrons by the MESSENGER spacecraft at 0.48 AU \citep{Feldman2010EvidenceSpectrometer, Lawrence2014DetectionOrigin}, but it does not align with the broader power-law spectra of albedo neutrons reported in ISS-RAD or modeled in past studies \citep{Clem2004NewSpectra, Wiegel2002SpectrometryLevel}.

Cosmic-ray albedo neutrons and ISS secondary neutrons are known to span a wide energy range, typically tens to hundreds of MeV. The high energy of these neutrons makes them difficult to thermalize within DANSON’s limited detector stack. Moreover, ISS-RAD measurements show that the differential flux of 2–4 MeV neutrons within the ISS is approximately 0.36 cm$^{-2}$s$^{-1}$ MeV$^{-1}$ \citep{Zeitlin2023ResultsDetector}, whereas the modeled fluence from DANSON exceeds this baseline by a significant margin in the zenith-facing direction.% The DANSON spectral profile resembles a narrow Gaussian centered at 1.5 MeV, unlike the power-law distribution (slope ~–0.83) typical of background flux inside the ISS.}
Together, the directional anisotropy, narrow spectral fit, and fluence comparison with established background levels indicate that the detected neutrons were not predominantly produced by spacecraft interactions or albedo processes.% Rather, the data support the interpretation of a predominantly solar-origin neutron flux.}

It is important to emphasize that the dominant zenith signal observed in DANSON cannot be attributed to spacecraft-induced secondary neutrons or albedo backgrounds. If such internal sources dominated, the nadir-facing thermoluminescent detectors (e.g., Layer 5) would be expected to register comparable or even higher signal levels due to neutron backscatter and structural reflection, as demonstrated in detailed albedo modeling studies \citep{Morris1995NeutronCOMPTEL, Morris1998ComptelOrbit, Selesnick2022ModelingProtons}. For example, Morris et al. reported nadir-to-zenith flux ratios of approximately 0.54 in their space-based neutron measurements, which indicate modest shielding effects. In contrast, DANSON recorded a nadir-to-zenith TL ratio of approximately 0.18 (which is evident in Figure 3 of the DANSON paper), representing a significantly more asymmetric flux distribution. This pronounced directional suppression, combined with the TL depth profile that matches MCNP6 simulations only in the 1 to 2 MeV energy range (peaking at $\sim$2.45 MeV), strongly disfavors internal secondary or lateral entry neutron scenarios. Although DANSON lacks imaging or time-of-flight resolution, its depth-resolved and directionally shielded design provides compelling evidence for an external directional component, consistent with a population of solar-origin neutrons.

%\section{Evaluation of Secondary Neutron Contributions from Local Proton Flux}
%To estimate the neutron yield from local secondary production, we used the GOES A4E channel, which captures protons with energies exceeding ~10 MeV. This channel represents the threshold for significant neutron-generating interactions with ISS structural materials such as aluminum and iron. While higher-energy protons in the A5E and A6E channels can contribute to neutron production, their flux during the DANSON observational period remained comparatively low. The contribution from A1E and A2E channels was neglected due to their sub-threshold energies, which are insufficient to produce 2–4 MeV secondary neutrons via nuclear spallation.

\section{Energy spectrum of solar neutrons}

\begin{figure*}
    \centering
    \includegraphics[width=\textwidth]{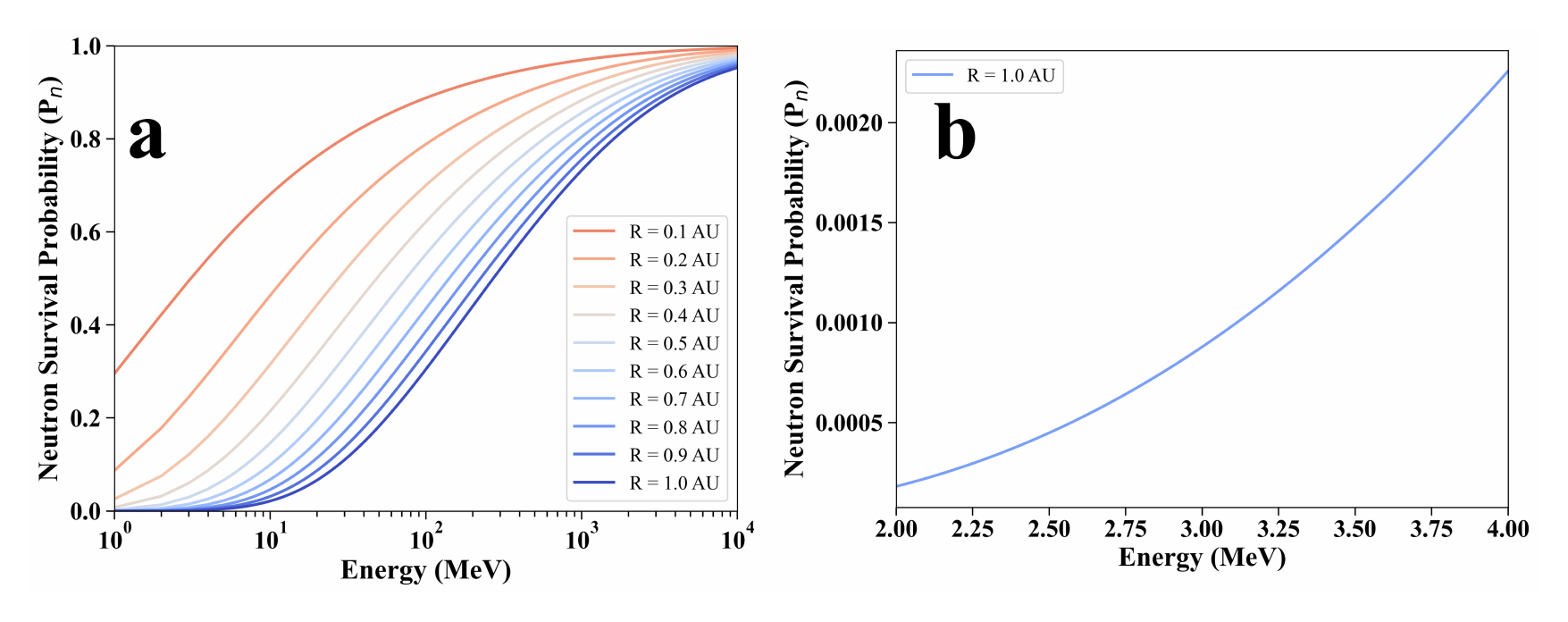}
    \caption{(a) The Solar neutron survival probability \(P_n\) at different distances from the Sun as a function of neutron energy, (b) Solar neutron survival probability \(P_n\) at 1AU distance from the Sun for the energy range 2.0 – 4.0 MeV. }
    \label{fig:probability}
\end{figure*}

The existing probabilistic model to estimate flux and fluences of Solar neutrons is connected to Solar flare events. The energy spectrum of solar neutrons associated with any solar flare is defined by,
\begin{equation}
    \varphi(E)dE = (R^{-2})Q(E,t)P_{n}(E)dE
\end{equation}
where \(R\) is the distance from the Sun, \(Q(E,t)\) the emissivity spectrum of neutrons emitted from Sun at time \(t\) and \(P_n(E)\) is the neutron survival probability that is the probability of the neutron of energy \(E\) will reach the distance \(R\) without decaying. \(P_n(E)\) is defined by the relativistic expression,

\begin{equation}
    P_n(E) = e^{-\frac{t'}{\tau}} = \exp \left[ \frac{-R}{\tau c \sqrt{[(E + mc^2)/mc^2] - 1}} \right]
\end{equation}
where \(t'\) is the time measured in the reference frame of the moving neutron, \(\tau\) is the mean lifetime (881.5 s) of a neutron and \(m\) the mass of a neutron (939.565560 MeV/c$^2$). Figure \ref{fig:probability}(a) depicts the variation of neutron survival probability \(P_n\) as a function of neutron energy \(E\) for different Heliocentric distances \(R\) and Figure \ref{fig:probability}(b) shows the same for the energy range 2 – 4 MeV at \(R\) = 1.0 AU. It is clear from the figure that the survival probability of neutrons having energies $<$ 10 MeV is very low. 

%%%%%%%%%%%%%%%%%%%%%%%%%%%%%%%%%%%%%%%%%%%%%%%%%%%%%%%%%%%%%%%%%%%%%%%%%%%%%%%%%%%%%%%
%									Figure - 1										  %
%%%%%%%%%%%%%%%%%%%%%%%%%%%%%%%%%%%%%%%%%%%%%%%%%%%%%%%%%%%%%%%%%%%%%%%%%%%%%%%%%%%%%%%
%%%%%%%%%%%%%%%%%%%%%%%%%%%%%%%%%%%%%%%%%%%%%%%%%%%%%%%%%%%%%%%%%%%%%%%%%%%%%%%%%%%%%%%

\begin{figure*}
    \centering
    \includegraphics[width=\textwidth]{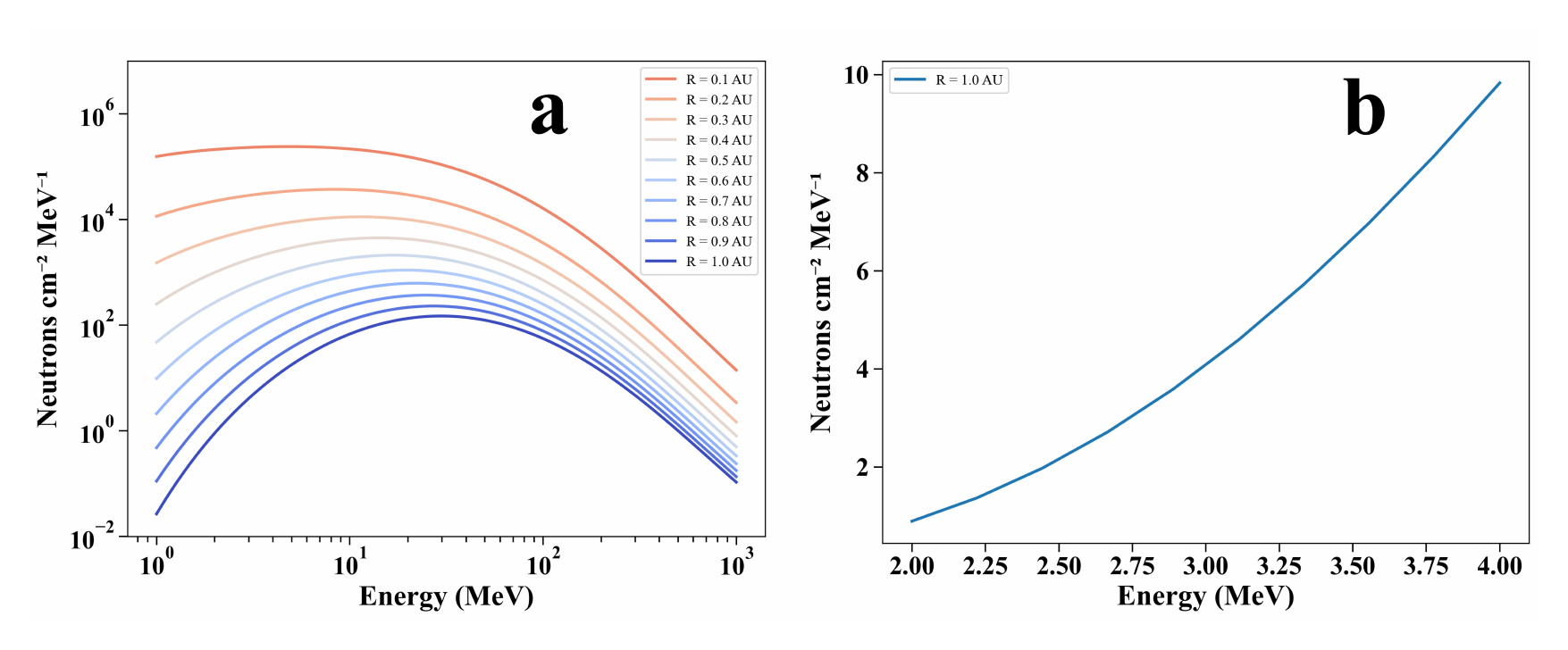}
    \caption{(a) The Energy spectra of solar neutrons at different heliocentric distances, (b) Energy spectra of solar neutrons at 1AU distance from the Sun in the energy range 2.0 – 4.0 MeV.}
    \label{fig:energy_spectrum}
\end{figure*}

The model algorithm used to determine the differential neutron emissivity \(Q(E,t)\) was developed by \cite{Hua1987SolarDistributions} while \cite{Murphy2007UsingLoop} updated the model to demonstrate elevated values of \(Q(E,t)\). \cite{Lario2012EstimationSun} used a 5th order polynomial fit normalized by the ~ 100 MeV neutron flux value of the largest Solar flare (4.3 × 10$^{28}$ neutrons MeV$^{-1}$ sr$^{-1}$) for the updated \(Q(E,t)\) model to determine the time-integrated energy spectrum of Solar neutrons \(\varphi(E)\) at different Heliocentric distances. 

\begin{align}
	\log_{10}(Q') &= 30.07 + 0.01 \log_{10}(E) - 0.39 \log_{10}^2(E) \notag \\
                  &\quad + 0.23 \log_{10}^3(E)
\end{align}

where \(E\) is in units of MeV and \(Q'\) is in units of neutrons MeV$^{-1}$ sr$^{-1}$.\\
Figure \ref{fig:energy_spectrum}(a) shows the energy spectra of Solar neutrons \(\varphi(E)\) for \(R\) = 0.1 – 1.0 AU considering obtained by Lario (2012). Figure \ref{fig:energy_spectrum}(b) shows the flux of 2 – 4 MeV neutrons at 1 AU.

%%%%%%%%%%%%%%%%%%%%%%%%%%%%%%%%%%%%%%%%%%%%%%%%%%%%%%%%%%%%%%%%%%%%%%%%%%%%%%%%%%%%%%%
%									Figure - 2										  %
%%%%%%%%%%%%%%%%%%%%%%%%%%%%%%%%%%%%%%%%%%%%%%%%%%%%%%%%%%%%%%%%%%%%%%%%%%%%%%%%%%%%%%%
%%%%%%%%%%%%%%%%%%%%%%%%%%%%%%%%%%%%%%%%%%%%%%%%%%%%%%%%%%%%%%%%%%%%%%%%%%%%%%%%%%%%%%%

%%%%%%%%%%%%%%%%%%%%%%%%%%%%%%%%%%%%%%%%%%%%%%%%%%%%%%%%%%%%%%%%%%%%%%%%%%%%%%%%%%%%%%%
%									Figure - 3										  %
%%%%%%%%%%%%%%%%%%%%%%%%%%%%%%%%%%%%%%%%%%%%%%%%%%%%%%%%%%%%%%%%%%%%%%%%%%%%%%%%%%%%%%%
\begin{figure*}
    \centering
    %\vspace{-1cm}
    \includegraphics[width=\textwidth]{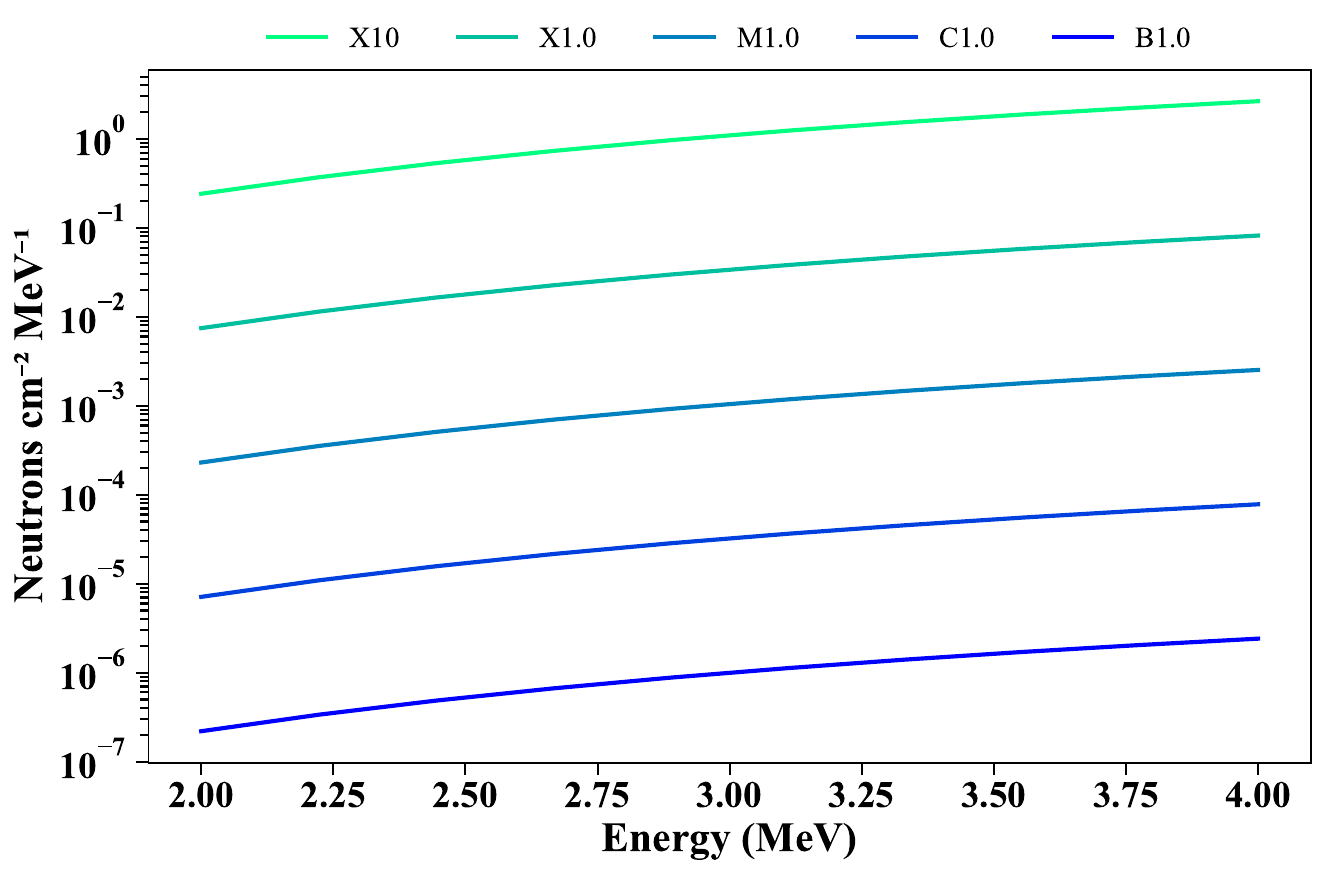}
    \caption{The Energy spectra of solar neutrons at 1AU distance scaled with different SXR flare classes.}
    \label{fig:spectra_SXR}
\end{figure*}
%%%%%%%%%%%%%%%%%%%%%%%%%%%%%%%%%%%%%%%%%%%%%%%%%%%%%%%%%%%%%%%%%%%%%%%%%%%%%%%%%%%%%%%

\cite{Lario2012EstimationSun} further scaled the \(Q(E,t)\) values according to a proportionality relationship that shows the correlation between the flux of neutrons at the solar surface and the soft X-ray (SXR) flux of the associated solar flares,
%\begin{equation}
%\begin{align}
%    \log_{10}(Q') &= 30.07 + 0.01 \log_{10}(E) - 0.39 \log_{10}^2(E) + 0.41 \log_{10}^3(E) \notag \\
%    &\quad - 0.30 \log_{10}^4(E) + 0.05 \log_{10}^5(E),
%\end{align}
%where \(E\) is in units of MeV and \(Q'\) is in units of neutrons MeV$^{-1}$ sr$^{-1}$.\\

\begin{equation}
    \log_{10}(Q'(E)) = 1.51\times log_{10}(SXR) + 3.96
\end{equation}
where SXR is the soft X-ray peak flux in W/m$^2$.

This proportionality relation is developed considering neutron observations for SXR values associated with class X flares. We used the same relation to scale the \(\varphi(E)\) values considering SXR values for flares of class X10, X1.0, M1.0, C1.0 and B1.0 for the neutron energy range 2 – 4 MeV as shown in Figure~\ref{fig:spectra_SXR}. It must be noted that during the DANSON experiment mostly class B and class C flares were detected. The respective flux for the class B and C flares ranges between 10$^{-7}$ and 10$^{-4}$ neutrons cm$^{-2}$ MeV$^{-1}$ for 2 – 4 MeV neutrons, which is incredibly low compared to that detected by DANSON. Hence, the low-energy neutrons may be majorly driven by some other phenomena and not only by solar flares or more likely, an accuerate estimate of the true low energy neuttron flux cannot be ascertained without better background subtraction of other particle interaction events from the detector(s).

\section{Scaling with CME width \(\alpha\)}

It has been established that impulsive solar energetic particles can lead to the production of low-energy neutrons. Observational studies suggest that impulsive solar energetic particles are often associated with narrow-width CMEs. To account for this effect, we introduce a CME width scaling factor to modify the neutron energy spectrum equation. The introduction of a CME width scaling factor in our model is based on the observed relationship between impulsive solar energetic particle events and narrow CMEs. Previous studies have established that impulsivesolar energetic particles are strongly associated with magnetic reconnection processes rather than large-scale shock-driven acceleration (Reames, 2014). Given that solar energetic particle acceleration and neutron production share similar underlying mechanisms, the inclusion of CME width as a modulating factor is well-supported. Furthermore, observational evidence suggests that narrow CMEs constrain energetic particle acceleration within a limited angular region, which provides justification for scaling neutron flux accordingly. Since the intensity of impulsive solar energetic particle events is inversely proportional to the CME width angle \( \alpha \), we assume that the neutron energy spectrum \( \varphi(E) \) follows a similar dependence:

\begin{equation}
    \phi(E) \propto \frac{1}{\alpha}
\end{equation}
The justification for this scaling relation can be derived from neutron production and escape probability arguments.

\subsection{Derivation of the CME Width Scaling Factor}

Neutrons are produced in solar flares andsolar energetic particleevents, where the CME width $\alpha$ determines the angular spread of the accelerated particles. If neutron production occurs uniformly within this confined region, the effective neutron emissivity per unit solid angle can be written as:
\begin{equation}
    Q_{\text{eff}}(E) = \frac{Q(E)}{\Omega_{\text{CME}}}
\end{equation}
where $Q(E)$ is the total neutron emissivity and $\Omega_{\text{CME}}$ is the solid angle subtended by the CME. The solid angle covered by a CME of width $\alpha$ is approximately:
\begin{equation}
    \Omega_{\text{CME}} \approx 2\pi \left( 1 - \cos\left(\frac{\alpha}{2}\right) \right)
\end{equation}
For moderate CME widths ($\alpha < 100^\circ$), using a Taylor series expansion of $\cos(x)$, we approximate:
\begin{equation}
    \Omega_{\text{CME}} \approx \pi \frac{\alpha^2}{4}
\end{equation}
In this study, we considered all CMEs with widths ranging from 20$^\circ$ to 80$^\circ$ and velocities exceeding 800 km/s, irrespective of whether they are associated with impulsive or gradualsolar energetic particleevents. While previous studies have found distinct differences in the solar energetic particle acceleration mechanisms between impulsive and gradual events, our approach focuses solely on the kinematic properties of CMEs rather than their specific solar energetic particle classification. As a result, the derived neutron flux scaling is applicable to a broad range of CME-drivensolar energetic particleevents. The approximation presented in Equation-9 is well-justified within this range, as the Taylor series expansion remains accurate for small to moderate angles. While this approximation may overestimate the solid angle for broader CMEs (130$^\circ$–180$^\circ$), such events are not considered in our analysis, making the current scaling relation appropriate for our study.
Since the observed neutron flux at a distance \(R\) from the Sun depends on the escaping neutron intensity, we write:
\begin{equation}
    \phi(E) = \frac{Q_{\text{eff}}(E)}{R^2} P_n(E)
\end{equation}
Substituting $Q_{\text{eff}}(E)$, we obtain:
\begin{equation}
    \phi(E) \approx \frac{Q(E) P_n(E)}{\Omega_{\text{CME}} R^2}
\end{equation}
Using the expression for $\Omega_{\text{CME}}$, this simplifies to:
\begin{equation}
    \phi(E) \propto \frac{4}{\pi \alpha^2}
\end{equation}
For moderate CME widths, the dependence approaches a near-linear relation, allowing us to approximate:
\begin{equation}
    \phi(E) \propto \frac{1}{\alpha}
\end{equation}

This inverse proportionality is consistent with empirical solar energetic particle observations \citep{Reames2014AbundanceEjections, Kahler2001CoronalEvents}, where the solar energetic particle fluence has been shown to decrease with the increasing CME width. Although the exact power-law dependence remains uncertain, empirical studies indicate that the solar energetic particle fluence is almost as strongly dependent on the CME width as on the CME speed \citep{Reames2014AbundanceEjections}. Observations further show that narrow CMEs tend to be more efficient in accelerating the solar energetic particles, and most events with widths below $50^\circ$ exhibit significantly higher fluences \citep{Reames2014AbundanceEjections}. This suggests that a simple inverse relation, rather than a strict quadratic dependence, better describes the neutron flux scaling with the CME width. Thus, including $1/\alpha$ as a scaling factor in our neutron flux model is justified.
Incorporating this into the standard neutron flux equation, the CME-width-dependent flux relation becomes:
%\begin{equation}
%    \phi(E)dE = \frac{360}{\alpha} (R^{-2}) Q(E,t) P_n(E) dE
%\end{equation}
%%%%%%%%%%%%%%%%%%%%%%%%%%%%%%

%%%%%%%%%%%%%%%%%%%%%%%%%%%%%%%
%\begin{equation}
%    \varphi(E) \propto \frac{k}{\alpha}
%\end{equation}

%Thus, the neutron flux per unit energy can be expressed as:

\begin{equation}
    \varphi(E) dE = \frac{k}{\alpha} (R^{-2}) Q(E, t) P_n(E) dE
\end{equation}
where, \( R = 1 \) AU is the Sun-Earth distance, \( Q(E, t) \) is the neutron escape spectrum and \( P_n(E) \) represents the neutron production probability.

To ensure consistency, we normalized the scaling factor by setting \( k \) to its maximum possible value when the CME width is \textbf{\( 360^\circ \)}. This yields the final form of the neutron flux equation:

\begin{equation}
    \varphi(E) dE = \frac{360}{\alpha} (R^{-2}) Q(E, t) P_n(E) dE
\end{equation}

\begin{figure}
    \centering
    %\vspace{-2cm}
    \includegraphics[width=\textwidth]{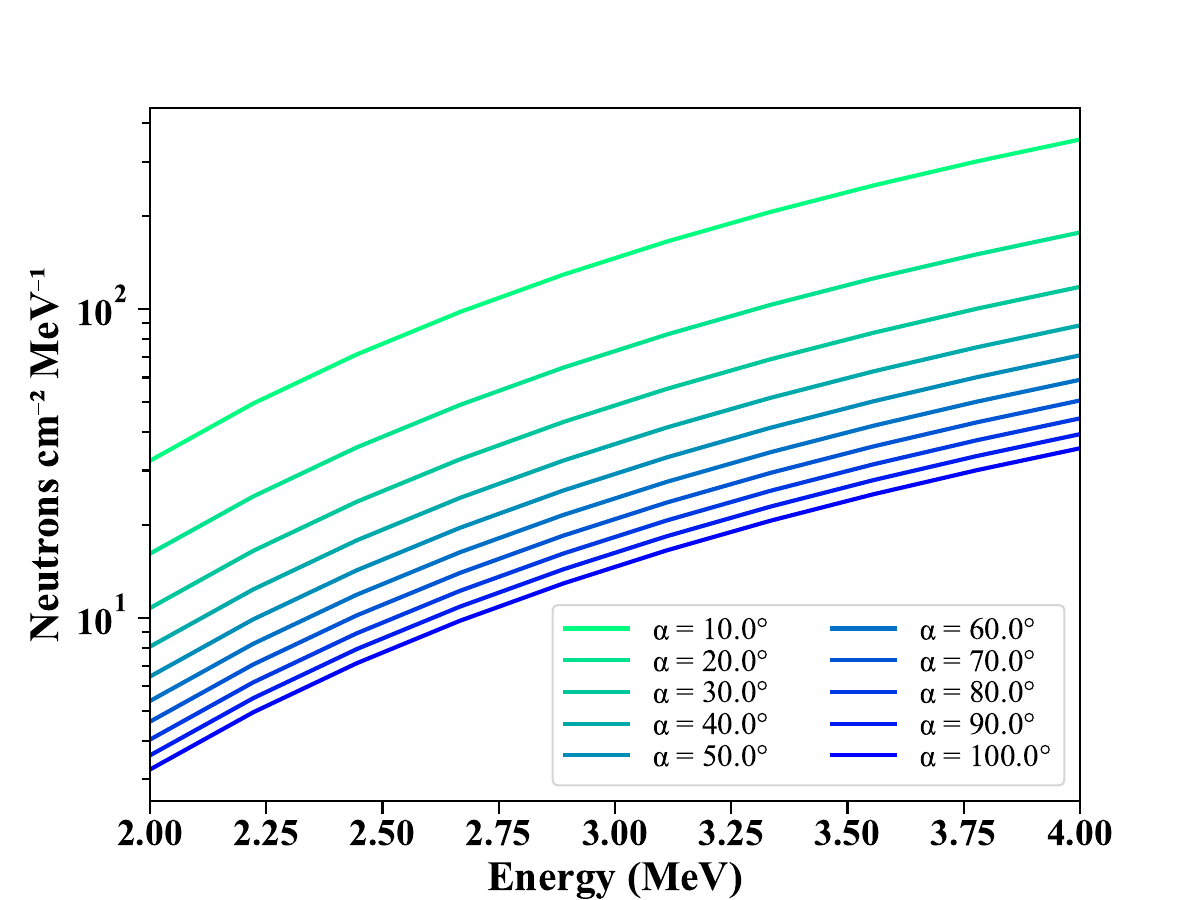}
    \caption{The Energy spectra of Solar neutrons at 1AU scaled with different CME widths.}
    \label{fig:spectra_CME}
\end{figure}

Using this equation, we analyzed the variation in neutron energy spectrum at a distance \(R\) = 1 AU for CME width values in the range \( 20^\circ - 80^\circ \) for the energy range of 2–4 MeV. Figure~\ref{fig:spectra_CME} shows the variation of \(\varphi(E)\) for the different values \(\alpha\) at 1 AU. Our results indicate that including CME width data in neutron flux modeling leads to a significant enhancement in the predicted neutron flux values, particularly for events associated with narrow CMEs. 

It is important to note that the current scaling relation does not explicitly account for the detailed angular distribution of neutron emission, which depends on the pitch-angle distribution of the accelerated ion population. Previous studies \citep{Hua1987SolarDistributions, Hua2002AngularLoops} have shown that neutron emission anisotropy can vary significantly depending on magnetic field geometry and scattering processes. While our model approximates the CME width as a proxy for angular confinement, a full treatment incorporating neutron pitch-angle transport would require more advanced simulation frameworks. Hence, the results presented here should be viewed as a first-order approximation pending such detailed modeling.

%%%%%%%%%%%%%%%%%%%%%%%%%%%%%%%%%%%%%%%%%%%%%%%%%%%%%%
%									Figure - 4										  %
%%%%%%%%%%%%%%%%%%%%%%%%%%%%%%%%%%%%%%%%%%%%%%%%%%%%%%
%%%%%%%%%%%%%%%%%%%%%%%%%%%%%%%%%%%%%%%%%%%%%%%%%%%%%%

\section{CME based low energy Solar neutron flux model}

From the previous section it is clear that for low-energy solar neutrons (2–4 MeV), the time-integrated energy spectrum shows relatively higher values when scaled using the CME width data. In this section, we determine the flux of low-energy solar neutrons by constraining the time-dependent neutron flux equation with CME width data. By incorporating the CME width \(\alpha\) into the time-dependent neutron flux equation, we aim to refine predictions of neutron flux at 1 AU and assess how constrained solar energetic particle acceleration impacts the observed neutron spectra. 

The time-dependent neutron flux at a distance \(R\) from the Sun is given by \cite{Hua1987SolarDistributions},

\begin{equation}
    F(t) = R^{-2}\times Q(E,t) \times \exp \left[ -\frac{\sqrt{t^2 - R^2/c^2}}{\tau}\right] 
\end{equation}
where \(F(t)\) represents the neutron flux, \(R\) is the distance in AU from the Sun, \(Q\) the neutron emissivity spectrum, \(t\) the time taken by a neutron of certain energy to reach the distance \(R\) and \(\tau\) = 881.5 s is the mean lifetime of the neutrons.

In the standard flux model, the total neutron flux received at Earth is obtained by integrating over all emission directions. If neutron emission is isotropic, the total solid angle is:

\begin{equation}
\Omega_{\text{total}} = 4\pi \text{ steradians}
\end{equation}
Thus, the direction-independent flux is given by:

\begin{equation}
F'(t) = \int_{\Omega} F(t) d\Omega = 4\pi \times F(t)
\end{equation}

In reality, solar neutrons are not always emitted isotropically. The CMEs can constrain the neutron-producing region to a limited angular width represented by the CME width, due to their role in modulating the distribution of accelerated particles \citep{Gopalswamy2014AnomalousImplications, Papaioannou2016SolarCharacteristics, Bronarska2018VeryParticles}. Hence, the emission solid angle becomes,
\begin{equation}
    \Omega_{\text{total}} \approx 4\pi \times \frac{\alpha}{360}
\end{equation}

As the flux should be scaled according to the limited angular width, we replace the factor \(4\pi\) in the equations (18) by 360/\(\alpha\):
\begin{equation}
    F'(t) = \frac{360}{\alpha} \times F(t)
\end{equation}

Substituting \(F(t)\) from Equation (16), the final CME-based neutron flux model equation becomes,

\begin{equation}
    F'(t) = \frac{360}{\alpha} \times R^{-2} \times Q(E,t) \times \exp \left[ -\frac{\sqrt{t^2 - R^2/c^2}}{\tau}\right]
\end{equation}

Finally, we use equations (21) and the observed CME width \((\alpha)\) data to estimate the low-energy solar neutron flux for the temporal domain of the DANSON experiment.

\section{Results}
\begin{figure}
    %\vspace{-5cm}
    \centering
    \includegraphics[width=\textwidth]{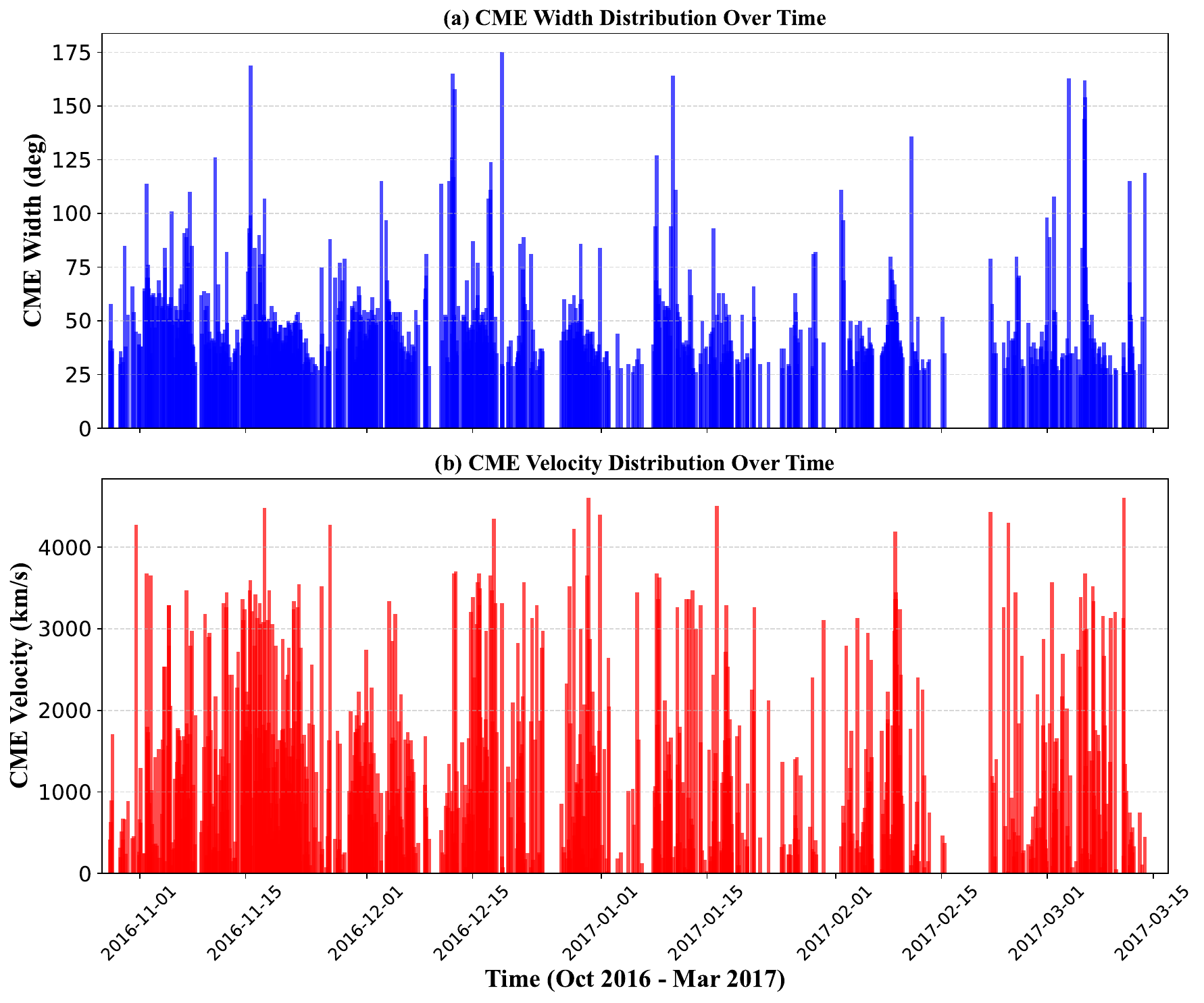}
    \caption{The Angular width (a) and velocity (b) of CMEs recorded by SEECHI COR2 in the duration of the DANSON experiment.}
    \label{fig:distri_CME_width}
\end{figure}

The model results of low-energy solar neutron flux for the CME observations are compared with the experimental observations in the temporal domain of the DANSON experiment (2016-10-27 to 2017-03-17). We extracted the respective width \((\alpha)\) and velocity \((v)\) data of all the CMEs in the datetime range 2016-10-28 01:24:00 UT to 2017-03-13 21:54:00 UT from the SECHHI COR2 database hosted on the SEEDS (Solar Eruptive Event Detection System) archives\footnote{\texttt{SEEDS:} \url{http://spaceweather.gmu.edu/seeds/secchi.php}}. Figure \ref{fig:distri_CME_width} shows the distribution of \(\alpha\) in panel (a) and \(v\) in panel (b) obtained from the SECHHI COR2 database in the duration of the DANSON experiment. 

\begin{figure}
    \centering
    %\vspace{-1cm}
    \includegraphics[width=\textwidth]{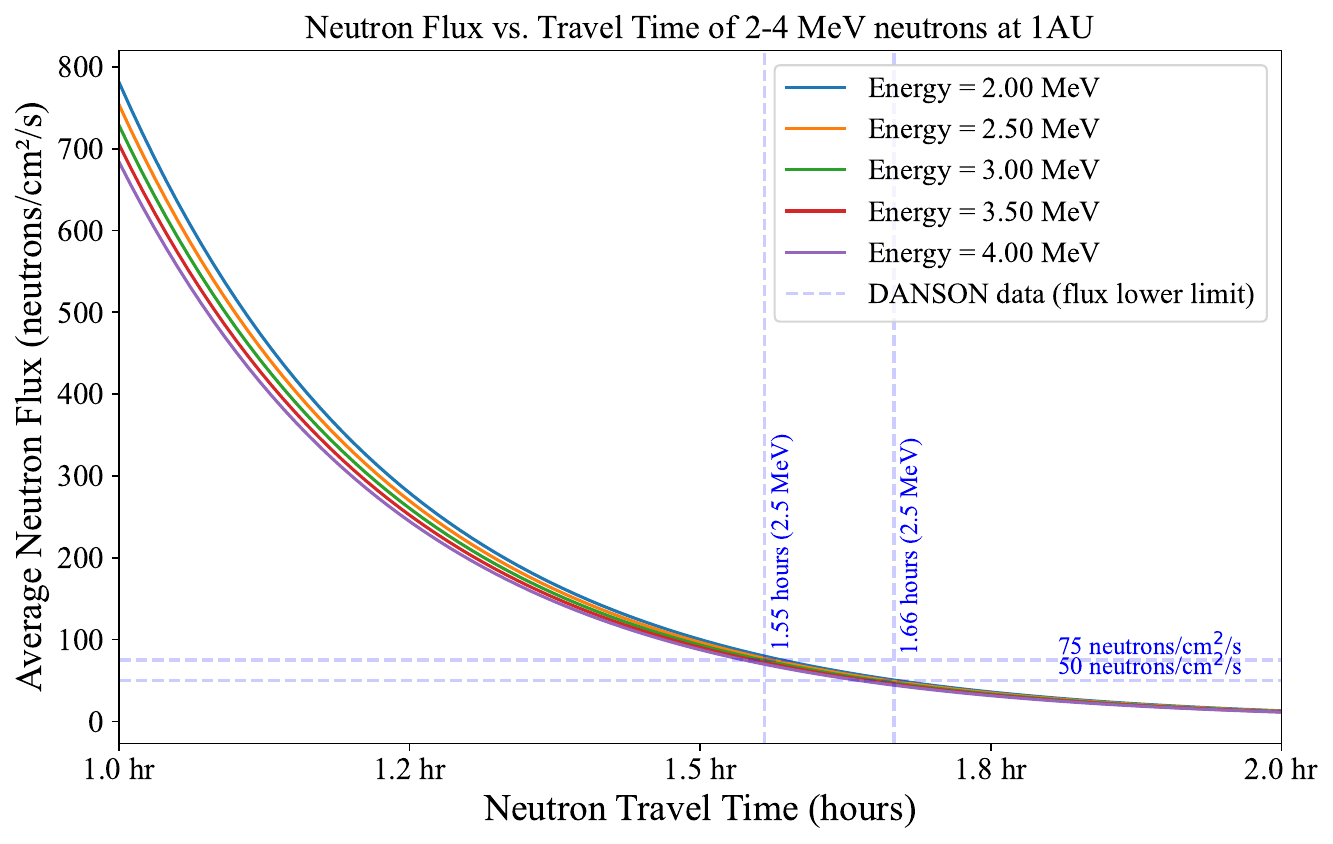}
    \caption{Distribution of average neutron flux \(F'_{avg}\) with different values of \(t\). Blue and the red lines shows the DANSON lower limit flux 50 and 75 neutrons/cm$^2$/s (dashed lines) and their respective neutron arrival times (solid lines) at 1AU.}
    \label{fig:flux_vs_time}
\end{figure}

It is important to note that not all CMEs may produce low-energy neutrons. Hence, to account for the neutron production from the CMEs we constrained the model for narrow-width \((20^{\circ} < \alpha < 80^{\circ})\) and faster CMEs \((v > 800\) km/s). We used this condition to determine the average flux \((F'_{avg})\) of low-energy neutrons in the energy range 2-4 MeV for the duration of the DANSON experiment. The variation of \(F'_{avg}(t)\) with respect to the neutron travel time \(t\) at 1 AU indicates the neutron flux reported (50-75 neutrons cm$^{-2}$ s$^{-1}$) from the DANSON experiment fits \(t\) = 1.55 to 1.66 hours according to the CME width-based neutron flux model (see Figure~\ref{fig:flux_vs_time}). A sharp decrease in modeled neutron flux is expected and is consistent with neutron decay theory, which emphasizes that only a tiny fraction of emitted neutrons actually reach Earth through these energies.

%%%%%%%%%%%%%%%%%%%%%%%%%%%%%%%%%%%%%%%%%%%%%%%%%%%%%%
%									Figure - 5										  %
%%%%%%%%%%%%%%%%%%%%%%%%%%%%%%%%%%%%%%%%%%%%%%%%%%%%%%
%%%%%%%%%%%%%%%%%%%%%%%%%%%%%%%%%%%%%%%%%%%%%%%%%%%%%%

%%%%%%%%%%%%%%%%%%%%%%%%%%%%%%%%%%%%%%%%%%%%%%%%%%%%%%
%									Figure - 6										  %
%%%%%%%%%%%%%%%%%%%%%%%%%%%%%%%%%%%%%%%%%%%%%%%%%%%%%%
%%%%%%%%%%%%%%%%%%%%%%%%%%%%%%%%%%%%%%%%%%%%%%%%%%%%%%

%%%%%%%%%%%%%%%%%%%%%%%%%%%%%%%%%%%%%%%%%%%%%%%%%%%%%%
%									Figure - 7										  %
%%%%%%%%%%%%%%%%%%%%%%%%%%%%%%%%%%%%%%%%%%%%%%%%%%%%%%
\begin{figure}
    \centering
    \includegraphics[width=\textwidth]{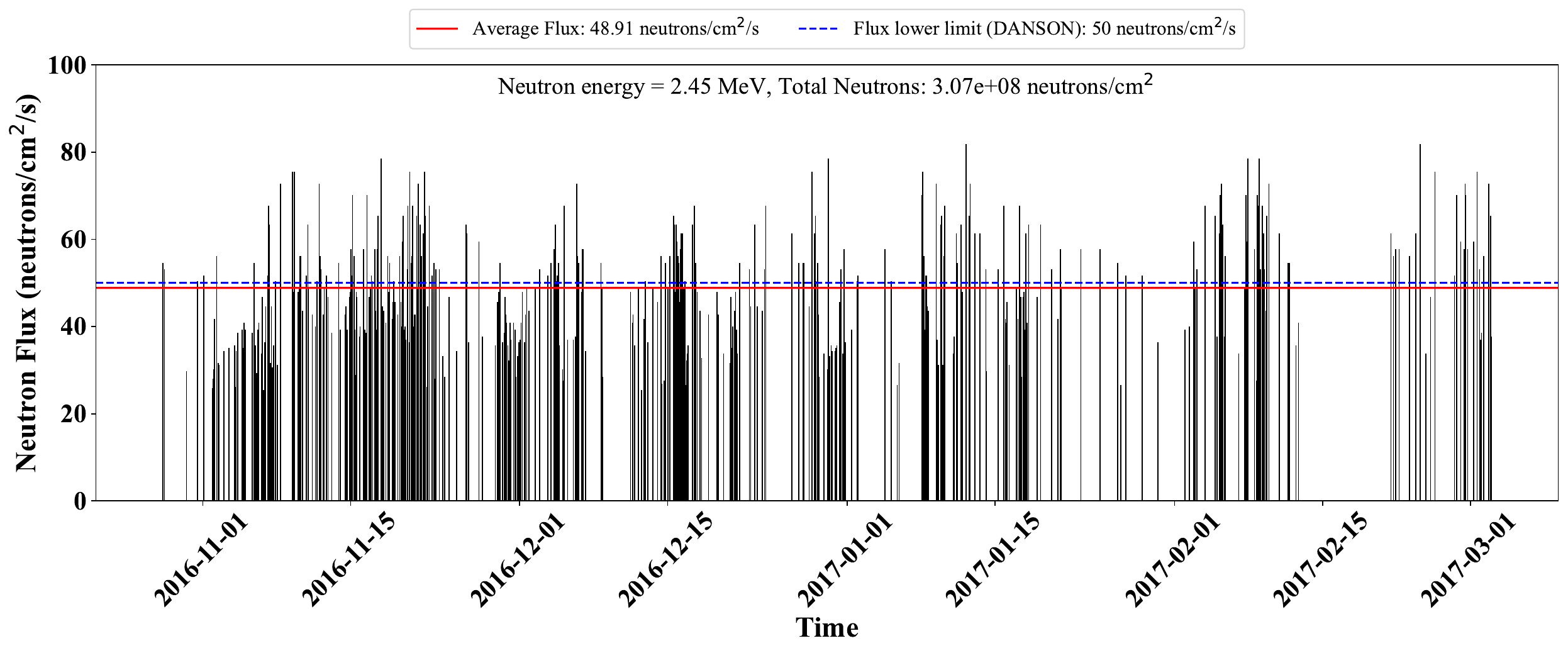}
    \caption{Modelled neutron flux (black) corresponding to the CMEs recorded in the duration of the DANSON experiment.}
    \label{fig:distri_flux}
\end{figure}
%%%%%%%%%%%%%%%%%%%%%%%%%%%%%%%%%%%%%%%%%%%%%%%%%%%%%%

To estimate the average and total flux of 2.45 MeV neutrons at 1AU during the duration of the DANSON experiment, we fixed the neutron travel time at \(t\) = 1.66 hrs = 6000 s and considered a flux time period range \(T\) = 4 - 6 hrs. Table 1 lists the total flux of 2.45 MeV neutrons at 1AU for the different values of \(T\) and compares them with the measured total flux (\(3 \times 10^9\) neutrons/cm$^2$) of 2-4 MeV neutrons obtained from the DANSON experiment. 
\begin{table}[h]
    \centering
    \begin{tabular}{lcc}
        \hline
        Flux Time Period (hours) & Total Flux (neutrons/cm$^{2}$) & DANSON total flux (\(3 \times 10^{9}\) neutrons/cm$^{2}$)\\
        \hline
        4 & \(3.07 \times 10^{8}\) & 10.23\%\\
        5 & \(3.84 \times 10^{8}\) & 12.8\% \\
        6 & \(4.61 \times 10^{8}\) & 15.4\% \\
        \hline
    \end{tabular}
    \caption{ Comparison between the total neutron flux predicted by the CME-width-based model and the neutron flux measured by the DANSON experiment for 2.45 MeV neutrons at 1 AU. Flux predictions are provided for different neutron emission duration periods (4 to 6 hours). The percentage column represents the fraction of the modeled neutron flux relative to DANSON’s measured total neutron flux (\(3 \times 10^9\) neutrons/cm$^2$)}
    \label{tab:flux_danson}
\end{table}

It is clear from the table that if the flux duration is at least 4 hours, the model generates a total flux of \(3.07 \times 10^8\) neutrons/cm$^2$ which is 10.23\% of the total flux obtained from the DANSON experiment. Also, the average flux obtained from this analysis is 48.91 neutrons cm$^{-2}$ s$^{-1}$ which falls close to the lower limit of the flux obtained from the DANSON experiment. Figure \ref{fig:distri_flux} shows the distribution of the model flux \((F'(t))\) of 2.45 MeV neutrons at 1 AU for flux time period \(T\) = 4 hours covering all CMEs with width \(20^{\circ} < \alpha < 80^{\circ}\) and velocity \(v > 800\) km/s in the temporal domain of the DANSON experiment.

\section{Discussion \& Conclusions}
%\subsection{Discussion}
This study investigates the role of coronal mass ejections (CMEs) in modulating low-energy solar neutron flux by introducing a CME-width scaling factor into existing neutron fluence models. By analyzing CME data from the SECHHI COR2 database during the DANSON experiment (2016-10-27 to 2017-03-17), we found that narrow-width ($20^\circ < \alpha < 80^\circ$) and fast ($v > 800$ km/s) CMEs play a significant role in neutron production. The revised model predicts a neutron flux of 48.91 neutrons cm$^{-2}$ s$^{-1}$ of 2.45 MeV at 1 AU closely matching the neutron flux reported by DANSON. Although, our estimates could be smaller, with a lower bound of about 5 neutrons cm$^{-2}$ s$^{-1}$ as evident from Figure~\ref{fig:spectra_CME}.

A key result of this study is that the estimated total flux of 2.45 MeV neutrons over a 4-hour period accounts for approximately 10.23\% of the total flux observed by DANSON. While, this might initially appear to be poor agreement (as noted at the outset), suitable background subtraction was performed for the solar neutron calorimeter \citep{Benker2019PossibleMaterials}. However, as the flux measurements lacked adequate background subtraction, over-counting is certain to occur. This suggests that CME-driven acceleration mechanisms significantly influence low-energy solar neutron production, challenging prior assumptions that low-energy neutrons should not survive transit to Earth due to rapid neutron decay. The findings further highlight the need to reconsider neutron transport and survival models for the 2–4 MeV energy range.

Despite the success of this approach, certain limitations remain. While the CME-width scaling factor effectively enhances neutron flux predictions, the assumption of an inverse dependence ($\phi(E) \propto 1/\alpha$) lacks direct observational verification from neutron measurements. Future observations are required to confirm this scaling relationship using dedicated neutron spectrometry.

Another critical aspect is the challenge of background subtraction in spaceborne neutron detection. While DANSON’s lithium tetraborate-based detection system improves neutron signal isolation, secondary neutron interactions within the spacecraft environment may introduce contamination. Future studies should incorporate Monte Carlo simulations (e.g., GEANT4, MCNP6) to refine neutron transport modeling and improve background corrections.

%\textbf{A useful benchmark in the study of spaceborne neutron detection is the COMPTEL instrument aboard the Compton Gamma Ray Observatory (CGRO), which was capable of imaging high-energy solar neutrons (>10 MeV) using time-of-flight (TOF) and double-scatter reconstruction techniques (Morris et al., 1995, 1998). However, COMPTEL's background studies revealed zenith-to-nadir neutron flux ratios of approximately 1.8–2.0, values consistent with known spacecraft shielding asymmetries and albedo effects. In contrast, DANSON, operating in the 2–4 MeV energy range, reported a significantly higher zenith-to-nadir ratio of ~3.0 after proton background subtraction via its ${}^{10}$B/${}^{11}$B-layered thermoluminescent design. This stronger directional asymmetry, combined with moderation profiles modeled by MCNP6 / GEANT4 that matched only in the 2 to 4 MeV range, cannot be easily explained by internal secondaries, albedo neutrons, or geometric shielding effects alone. While DANSON lacks imaging capability, its passive long-duration integration and neutron-specific detection methodology offer superior sensitivity to persistent low-energy directional neutron fluxes — a domain where COMPTEL was not optimized. This enhanced directional response reinforces the argument that DANSON observed a statistically significant external neutron signal in a previously underexplored energy regime. }

A valuable benchmark in spaceborne neutron detection is the COMPTEL instrument aboard the Compton Gamma Ray Observatory (CGRO), which successfully imaged high-energy solar neutrons ($>$10~MeV) using time-of-flight (TOF) and double-scatter reconstruction techniques \citep{Morris1995NeutronCOMPTEL, Morris1998ComptelOrbit}. Background studies from COMPTEL revealed nadir-to-zenith neutron flux ratios of approximately 0.54 and 0.7, consistent with expectations from spacecraft shielding asymmetries and terrestrial albedo effects. In contrast, DANSON, which operated in the lower 2–4~MeV energy range, reported a significantly lower nadir-to-zenith TL signal ratio of $\sim$0.18 after proton background subtraction, leveraging its ${}^{10}$B/${}^{11}$B-layered thermoluminescent detector configuration. This enhanced directional asymmetry cannot be readily explained by internal secondary neutrons, geometric shielding, or known albedo distributions. Although DANSON lacked the directional imaging capabilities of COMPTEL, its long-duration passive integration and neutron-specific design made it uniquely sensitive to persistent low-energy directional neutron fluxes, a spectral domain outside the optimized range of COMPTEL. 

The strength of DANSON's signal is further supported by the shape of its thermoluminescent response: after background subtraction, the signal peaked at Layer~2 with clear suppression at the nadir-facing Layer~5. This depth-dependent distribution was best matched by MCNP6 simulations for neutron energies between 2 and 4~MeV \citep{Benker2019PossibleMaterials}. Simulations at lower energies (e.g., 0.8~MeV) resulted in capture profiles that were too shallow, while higher-energy neutrons (10~MeV) produced overly uniform distributions. Together, directional asymmetry and spectral selectivity reinforce the plausibility that DANSON detected a statistically significant, low-energy neutron signal of external, potentially solar origin, consistent with the modeled flux used in this study.

This study underscores the need for further observational campaigns with improved neutron detectors. While current and upcoming solar-focused missions such as Parker Solar Probe, Solar Orbiter, Aditya-L1, PUNCH (Polarimeter to Unify the Corona and Heliosphere), and the proposed Lagrange L5 mission are not explicitly designed for low-energy neutron detection, they offer complementary measurements of energetic particles, gamma rays, and CME dynamics. These data may help constrain the acceleration environments and seed populations associated with neutron-producing processes. Future progress in validating CME-based neutron flux models will require dedicated instrumentation capable of directional neutron detection and precise energy discrimination in the 1–10 MeV range.

\section*{Acknowledgments}
The authors acknowledge the SEEDS (Solar Eruptive Event Detection System) archives for providing access to the SECHHI COR2 database, which was vital for the CME data used in this study. The authors thank the teams responsible for the DANSON experiment aboard the International Space Station for publicly providing neutron flux measurements. This work was partly funded by the University of Nebraska-Lincoln and the University of Nebraska Grand Challenges catalyst award entitled Quantum Approaches addressing Global Threats.

%%% BIBLIOGRAPHY %%%%%%%%%%%%%%%%%%%%%%%%%%%%%%%%%%%%%%%%%%%%%%%%%%%%%%%%%%%

%% Bibliography
%% Author year style
\bibliographystyle{chicago}
\biboptions{authoryear}
\bibliography{ref_low_energy_neutrons}

\end{document}